\documentclass[twocolumn,showpacs,preprintnumbers,amsmath,amssymb]{revtex4}

\usepackage{graphicx}% Include figure files
\usepackage{dcolumn}% Align table columns on decimal point
\usepackage{bm}% bold math

\begin{document}

\preprint{APS/123-QED}

\title{A Toy Model for the Magnetic Connection\\between a Black Hole and
a Disk}

\author{Li-Xin Li}
  \altaffiliation[Current address: ]{Harvard-Smithsonian Center for 
  Astrophysics, 60 Garden St., MS-51, Cambridge, MA 02138}
  \email{lli@cfa.harvard.edu}
  \affiliation{Department of
  Astrophysical Sciences, Princeton University, Princeton, NJ 08544}

\date{November 28, 2001}

\begin{abstract}
A magnetic field connecting a Kerr black hole to a disk rotating around
it can extract energy and angular momentum from the black hole and transfer
them to the disk if the black hole rotates faster than the disk. The energy 
can be dissipated and radiated away by the disk, which makes the disk
shine without the need of accretion. In this paper we present a toy model
for the magnetic connection: a single electric current flowing around a Kerr
black hole in the equatorial plane generates a poloidal magnetic field which 
connects the black hole to the disk. The rotation of the black hole relative
to the disk generates an electromotive force which in turn generates a
poloidal electric current flowing through the black hole and the disk and
produces a power on the disk. We will consider two cases: (1) The toroidal 
current flows on the inner boundary of the disk, which generates a poloidal 
magnetic field connecting the horizon of the black hole to a region of the
disk {\it beyond} the inner boundary; (2) The toroidal current flows on a
circle inside the inner boundary of the disk but outside the horizon of the
black hole, which generates a poloidal magnetic field connecting a portion
of the horizon of the black hole to the {\it whole} disk. We will calculate
the power produced by the magnetic connection and the resulting radiation
flux of the disk in the absence of accretion, and compare them with that
produced by accretion.
\end{abstract}


\maketitle

%\section 1
\section{\label{sec1}Introduction}

A magnetic field connecting a Kerr black hole to a disk rotating around it
can transfer energy and angular momentum between the black hole and the disk
(\cite{bla99,li00a,li00b,li00c} and references therein). When the black hole
rotates faster than the disk, energy and angular momentum are extracted from
the black hole and transfered to the disk. If the magnetic field is weak, the
energy transfered to the disk by the magnetic connection can be dissipated
and radiated away by the disk \cite{li00a,li00b,li00c}. Thus, in the
presence of magnetic connection, the rotational energy of the black hole
provides an energy source for disk radiation in addition to disk accretion
\cite{nov73,pag74,tho74}. In fact, a disk magnetically coupled to a
rapidly rotating Kerr black hole can radiate without the need of accretion
\cite{li00a,li00b,li00c}. In such a case, all the power of the disk comes from
the rotational energy of the black hole.

Due to the lack of knowledge about the topology and distribution of the
magnetic field connecting the black hole to the disk, a detailed model of the
magnetic connection between a black hole and a disk does not exist yet, except 
the simplest and {\it ad hoc} model where the magnetic field is assumed to 
connect the black hole to the disk close to the inner boundary 
\cite{li00a,li00b}, or at a single radius beyond
the inner boundary \cite{li00c}. In this paper, we consider a toy model for
the magnetic connection: A single toroidal electric current flows around a
Kerr black hole in the equatorial plane which is coincident with the plane of
a thin Keplerian disk rotating around the black hole. The toroidal
current generates a poloidal magnetic field which connects the black hole to
the disk. Due to the relative rotation between the black hole and the disk and
the fact that the horizon of a Kerr black hole is a conductor with a finite
electric resistivity \cite{zna78,dam78,car79} and a plasma disk is a perfect
conductor, an electromotive forces (EMF) is induced which in turn generates a
poloidal electric current flowing through the black hole and the disk. The
poloidal current produces a torque and a power on the disk, which makes the
disk shine without the need of accretion. We will calculate the power and the 
resulting radiation flux of the disk with the assumption that there is no 
accretion and the configuration is steady and axisymmetric, and
compare the results with that of a standard accretion disk.

A similar model has been used by Li \cite{li00d} to test the efficiency of the
Blandford-Znajek mechanism \cite{bla77,mac82,phi83}. By assuming a toroidal
current continuously distributed in a thin Keplerian disk, Li \cite{li00d} has
shown that the power of the disk always dominates over the power of the black
hole, which confirms the early speculation of Blandford and Znajek \cite{bla77}
and the recent arguments of Livio, Ogilvie, and Pringle \cite{liv99}. In this
paper we will find that, an electric current flowing around a black hole in the 
equatorial plane naturally generates a magnetic field connecting the black hole 
to its disk. This implies that the existence of magnetic connection between a
black hole and a disk is a very natural assumption in view of physics.
Essentially the magnetic connection between a black hole and a disk is a variant
of the Blandford-Znajek mechanism. However, it is simpler and better defined 
than the Blandford-Znajek mechanism since a unknown remote load is not required
\cite{li00c}. Though the model presented in this paper is so simple that it
cannot be directly applied to real astronomy, it helps us in understanding
some physical aspects of the magnetic connection.

Throughout the paper we use the geometrized units $G = c = 1$ and the 
Boyer-Lindquist coordinates $(t,r,\theta,\phi)$ \cite{mis73,wal84}.

%\section 2
\section{The magnetic flux and the magnetic connection}

Assuming a neutral toroidal electric current $I$ flows on a circle of $r =
r^\prime$ in the equatorial plane of a Kerr black hole, then the four-current 
vector is
\begin{eqnarray}
    J^a = \frac{I}{r}\left(\frac{\Delta}{A}\right)^{1/2} \left(\frac{
          \partial}{\partial\phi}\right)^a \, \delta \left(r - r^\prime
          \right)\delta (\cos\theta) \,,
    \label{fcur}
\end{eqnarray}
where $\Delta \equiv r^2-2Mr+a^2$, $A \equiv (r^2+a^2)^2-\Delta a^2 \sin^2
\theta$, $M$ is the mass of the black hole, $M a$ is the angular momentum of
the black hole ($a^2\leq M^2$), and $\delta (x)$ is the Dirac $\delta$-function. 
The magnetic flux through a surface bounded by a circle with $r={\rm constant}$
and $\theta = {\rm constant}$, produced by such an electric current, is
\begin{eqnarray}
    \Psi(r,\theta; r^\prime) = 2\pi A_\phi(r,\theta; r^\prime) \,,
    \label{flux}
\end{eqnarray}
where $A_\phi$\,, the toroidal component of the electric vector potential, is
given by \cite{zna78a,lin79}
\begin{widetext}
\begin{eqnarray}
    A_\phi &=& 2\sum_{l=1}^\infty
    \left\{\alpha_l^r\left[ra\sin^2\theta\frac{\Delta}{\Sigma}\frac{1}
    {\sqrt{M^2-a^2}}P_l^\prime(u)P_l(\cos\theta)-a\sin^2\theta\cos\theta
    \frac{r^2+a^2}{\Sigma}P_l(u)P_l^\prime(\cos\theta)\right]\right.
    \nonumber\\
    &&+\alpha_l^i\left[-a^2\sin^2\theta\cos\theta\frac{\Delta}{\Sigma}
    \frac{1}{\sqrt{M^2-a^2}}P_l^\prime(u)P_l(\cos\theta)-r\sin^2\theta
    \frac{r^2+a^2}{\Sigma}P_l(u)P_l^\prime(\cos\theta)\right.\nonumber\\
    &&\left.\left.+\frac{\Delta\sin^2\theta}{l(l+1)}\frac{1}{\sqrt{
    M^2-a^2}}P_l^\prime(u)P_l^\prime(\cos\theta)\right]\right\}\nonumber\\
    &&+2\sum_{l=1}^\infty\left\{\beta_l^r\left[ra\sin^2\theta\frac{\Delta}
    {\Sigma}\frac{1}{\sqrt{M^2-a^2}}Q_l^\prime(u)P_l(\cos\theta)-a\sin^2\theta
    \cos\theta\frac{r^2+a^2}{\Sigma}Q_l(u)P_l^\prime(\cos\theta)\right]
    \right.\nonumber\\
    &&+\beta_l^i\left[-a^2\sin^2\theta\cos\theta\frac{\Delta}{\Sigma}\frac{1}
    {\sqrt{M^2-a^2}}Q_l^\prime(u)P_l(\cos\theta)-r\sin^2\theta\frac{r^2+a^2}
    {\Sigma}Q_l(u)P_l^\prime(\cos\theta)\right.\nonumber\\
    &&\left.\left.+\frac{\Delta\sin^2\theta}{l(l+1)}\frac{1}{\sqrt{
    M^2-a^2}}Q_l^\prime(u)P_l^\prime(\cos\theta)\right]\right\},
    \label{vec0}
\end{eqnarray}
\end{widetext}
where $u \equiv (r-M)/\sqrt{M^2-a^2}$, $P_l(z)$ and $Q_l(z)$ are Legendre
functions, $P_l^\prime(z) \equiv dP_l(z)/dz$, $Q_l^\prime(z) \equiv dQ_l(z)
/dz$, and the coefficients $\alpha_l^r$, $\alpha_l^i$, $\beta_l^r$, and
$\beta_l^i$ are respectively \\
(1) for $r<r^\prime$, $\beta_l^r=\beta_l^i=0$ for all $l$; but
\begin{eqnarray}
    \alpha_l^r \equiv \frac{\pi(2l+1) I}{l(l+1)(M^2-a^2)}\left(\frac{
        \Delta^\prime}{A^\prime}\right)^{1/2}\Delta^\prime a P_l(0)
        Q_l^\prime(u^\prime),
    \label{vec1}
\end{eqnarray}
\begin{eqnarray}
    \alpha_l^i &\equiv& \frac{\pi (2l+1) I}{l(l+1)\sqrt{M^2-a^2}}
        \left(\frac{\Delta^\prime}{A^\prime}\right)^{1/2}
        \left[-({r^\prime}^2+a^2)P_l^\prime(0)\right.\nonumber\\
        &&\left.\times Q_l(u^\prime)+\frac{r^\prime\Delta^\prime}{l(l+1)}
        \frac{1}{\sqrt{M^2-a^2}}P_l^\prime(0)Q_l^\prime(u^\prime)\right];
    \label{vec2}
\end{eqnarray}
(2) for $r>r^\prime$, $\alpha_l^r=\alpha_l^i=0$ for all $l$; but
\begin{eqnarray}
    \beta_l^r \equiv \frac{\pi (2l+1) I}{l(l+1)(M^2-a^2)}\left(\frac{
        \Delta^\prime}{A^\prime}\right)^{1/2}\Delta^\prime a P_l(0)
        P_l^\prime(u^\prime),
    \label{vec3}
\end{eqnarray}
\begin{eqnarray}
    \beta_l^i &\equiv& \frac{\pi (2l+1) I}{l(l+1)\sqrt{M^2-a^2}}
        \left(\frac{\Delta^\prime}{A^\prime}\right)^{1/2}
        \left[-({r^\prime}^2+a^2)P_l^\prime(0)\right.\nonumber\\
        &&\left.\times P_l(u^\prime)+\frac{r^\prime\Delta^\prime}{l(l+1)}
        \frac{1}{\sqrt{M^2-a^2}} P_l^\prime(0)P_l^\prime(u^\prime)\right];
    \label{vec4}
\end{eqnarray}
where $\Delta^\prime=\Delta(r=r^\prime)$ and $A^\prime=A(r=r^\prime,\theta=
\pi/2)$.

The magnetic flux through the hemi-sphere of the horizon of the black hole is
$\Psi(r_H, \pi/2; r^\prime)$, where $r_H = M + \sqrt{M^2-a^2}$ is the radius of 
the horizon. The magnetic flux through a surface bounded by a circle $r = {\rm
constant}$ in the equatorial plane is $\Psi(r, \pi/2; r^\prime)$. The magnetic 
field lines generated by the toroidal current are closed loops around the circle
$r = r^\prime$, except the field line along the axis $\theta = 0$. Consider
the field lines leaving the horizon at a polar angle $\theta_1$, they will
intersect the equatorial plane at a circle of radius $r_1$ where $r_1$ is given
by $\Psi(r_1,\pi/2; r^\prime) = \Psi(r_H, \theta_1; r^\prime)$. Clearly, if the
inner edge of the disk lying in the equatorial plane is inside the circle $r = 
r_1$ and the outer edge of the disk is outside the circle $r = r_1$, a magnetic 
connection must exist between the black hole and the disk.

Let us consider a thin Keplerian disk in the equatorial plane of a Kerr black
hole \cite{nov73,pag74,tho74}. The angular velocity of the disk is
\begin{eqnarray}
    \Omega_D(r) = \left(\frac{M}{r^3}\right)^{1/2}\frac{1}{1+a\left(\frac{M}
             {r^3} \right)^{1/2}} \,.
    \label{wd}
\end{eqnarray}
The outer edge of the disk is at $r = \infty$. The inner edge of the disk is
at the marginally stable orbit
\begin{eqnarray}
    r_{ms} = M\left\{3+z_2-\left[(3-z_1)(3+z_1+2z_2)\right]^{1/2}
             \right\}\,
    \label{rms}
\end{eqnarray}
where
\begin{subequations}
\begin{eqnarray}
    z_1 &\equiv& 1 + \left(1 - \frac{a^2}{M^2}\right)^{1/3} \nonumber\\
             &&\times\left[\left(1 + \frac{a}{M}\right)^{1/3}
             + \left(1 - \frac{a}{M}\right)^{1/3}\right] \,,\\
    z_2 &\equiv& \left(z_1^2 + 3 \frac{a^2}{M^2}\right)^{1/2} \,.
\end{eqnarray}
\end{subequations}

\begin{figure}
\includegraphics[width=8.5cm]{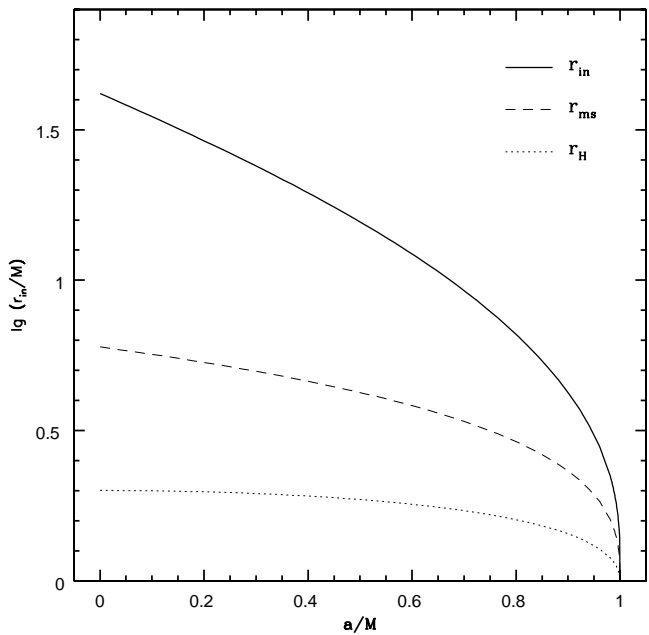}
\caption{\label{fig1} The radius where the innermost flux surface intersects 
the disk as a function of the spin of the black hole (the solid line): the
toroidal electric current flows on the marginally stable orbit $r^\prime =
r_{ms}$. For comparison, the inner edge of the disk (the dashed line) and the
horizon of the black hole (the dotted line) are also shown.}
\end{figure}

\begin{figure}
\includegraphics[width=8.5cm]{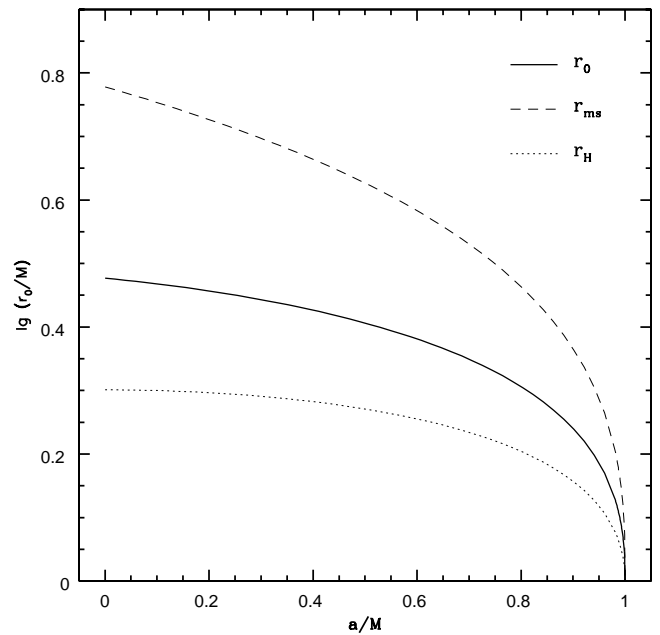}
\caption{\label{fig2} The radius where if the toroidal current is put then 
the generated magnetic field connects the portions of the horizon of the black
hole with $0<\theta < \pi/3$ and $2\pi/3 <\theta < \pi$ to the whole disk from 
$r = r_{ms}$ to $r =
\infty$ (the solid line). For comparison, the inner edge of the disk (the 
dashed line) and the horizon of the black hole (the dotted line) are also 
shown.}
\end{figure}

Due to the existence of a gap between the horizon of the black hole and the
inner boundary of the the disk, not all magnetic field lines connect the black
hole to the disk. If we define a flux surface to be a surface of constant
magnetic flux, there is an {\it innermost
flux surface} which connects the black hole to the disk. Assuming the toroidal
electric current is located at the inner boundary of the disk: $r^\prime =
r_{ms}$, then the innermost flux surface touches the horizon of the black
hole at $\theta = \pi/2$ and intersects the disk at a circle whose radius
$r_{in}$ is determined by
\begin{eqnarray}
    \Psi\left(r_{in},\frac{\pi}{2}; r_{ms}\right) =
         \Psi\left(r_H,\frac{\pi}{2}; r_{ms}\right) \,,
	 \hspace{0.3 cm} r_{in} > r_{ms} \,.
    \label{rin}
\end{eqnarray}
Since the magnetic field lines in the gap between the black hole and the disk
must close themselves on the disk, $r_{in}$ must be greater than $r_{ms}$.
Clearly, $r_{in}/M$ depends only on $a/M$ when $r^\prime = r_{ms}$ is fixed.
In Fig.~\ref{fig1} we plot $r_{in}/M$ as a function of $a/M$. For comparison,
we also plot $r_{ms}/M$ and $r_H/M$ in Fig.~\ref{fig1}. We see that, as $a/M$
increases, $r_{in}$ gets closer to $r_{ms}$. But, for $a/M < 1$, we always
have $r_{in} > r_{ms}$. The magnetic field lines between $r_{ms}$ and $r_{in}$
close themselves in the gap region between $r_H$ and $r_{ms}$, they do not
touch the black hole horizon.

In order for the magnetic field to connect the black hole to the whole disk
from $r = r_{ms}$ to $r = \infty$, the toroidal current must be inside the 
inner edge of the disk but outside the horizon of the black hole. Assume
$r^\prime = r_0$ when the inner most flux surface touches the disk at $r =
r_{ms}$ and the horizon of the black hole at $\theta = \pi/3$ (and $\theta
= 2\pi/3$ due to the reflection symmetry about the equatorial plane). Then, 
$r_0$ is determined by
\begin{eqnarray}
    \Psi\left(r_{ms},\frac{\pi}{2}; r_0\right) =
         \Psi\left(r_H,\frac{\pi}{3}; r_0\right) \,,
	 \hspace{0.3 cm} r_H < r_0 < r_{ms} \,.
    \label{r0}
\end{eqnarray}
Like $r_{in}/M$, $r_0/M$ also depends only on $a/M$. In Fig.~\ref{fig2} we plot
$r_0/M$ as a function of $a/M$. For comparison, we also plot $r_{ms}/M$ and 
$r_H/M$ in Fig.~\ref{fig2}.

\begin{figure}
\includegraphics[width=8.5cm]{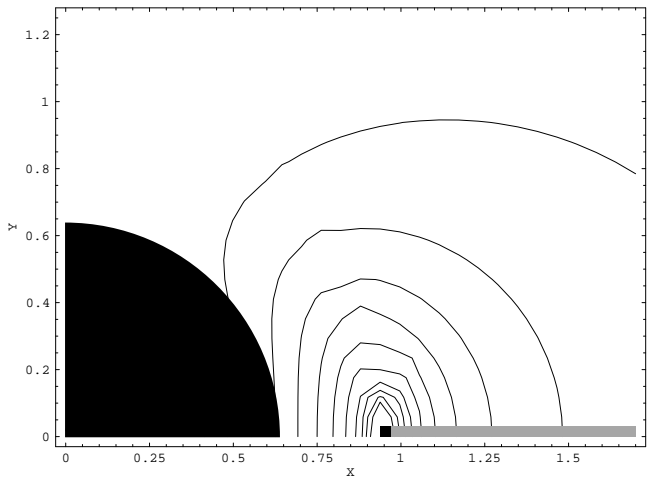}
\caption{\label{fig3} The topology of the magnetic field produced by a toroidal
electric current flowing on the marginally stable orbit in the equatorial plane
of a Kerr black hole of mass $M$ and specific angular momentum $a = 0.99 M$.
The horizontal axis is $X \equiv (r/M) \cos\theta - 0.5$, the vertical axis is 
$Y\equiv (r/M) \sin\theta - 0.5$. The black disk represents the inside of the 
black
hole, whose boundary is the horizon of the black hole. The gray line lying on
the X-axis is the Keplerian disk. The black dot represents the inner edge of
the disk which is at the marginally stable orbit. The solid curves represent
the magnetic field lines (tangent to the flux surfaces) generated by the
toroidal current. The innermost flux surface touches the horizon at $\theta =
0$ and touches the disk at a radius beyond the inner boundary.}
\end{figure}

\begin{figure}
\includegraphics[width=8.5cm]{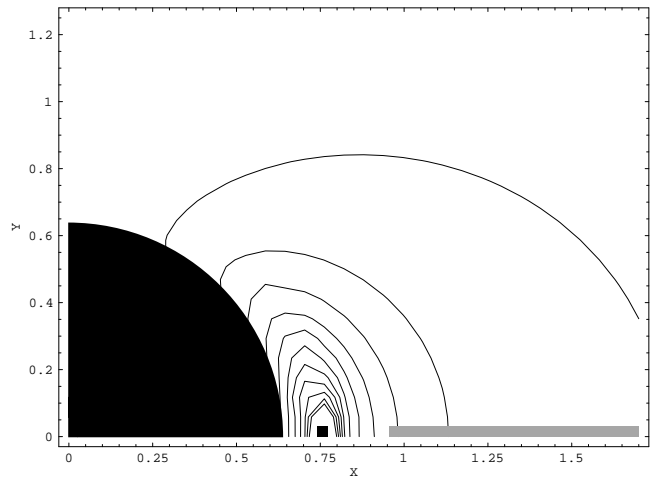}
\caption{\label{fig4} The topology of the magnetic field produced by a toroidal
electric current flowing on a circle in the equatorial plane. The model is the 
same as that in Fig.~\ref{fig1} except that now the electric current is inside
the inner boundary of the disk. The innermost flux surface touches the horizon 
of the black hole at $\theta = \pi/3$ (and $\theta = 2\pi/3$) and touches the 
disk at the inner boundary.}
\end{figure}

Examples of the topological structure of the magnetic field produced by a
toroidal electric current flowing in the equatorial plane of a Kerr black hole
are shown in Fig.~\ref{fig3} and Fig.~\ref{fig4}.

%\section 3
\section{The power produced by the magnetic connection}
\label{sec3}

In the presence of a magnetic field a Kerr black hole behaves like a conductor 
with a surface resistivity $R_H = 4\pi \approx 377$ Ohms 
\cite{zna78,dam78,car79}. A plasma disk can be treated as a perfect conductor 
with a vanishing resistivity. So, the rotation of the black hole and the disk 
induces an EMF on the horizon of the black hole and an EMF on the disk
\cite{mac82,li00d}. For a bunch of magnetic field lines connecting the black
hole to the disk within radii $r$ --- $r + \Delta r$, where $\Delta r \ll r$,
the induced EMFs on the horizon of the black hole and the disk are respectively
\begin{eqnarray}
    \Delta{\cal E}_H = \frac{1}{2\pi}\Omega_H \Delta\Psi \,, \hspace{0.5cm}
    \Delta{\cal E}_D = -\frac{1}{2\pi}\Omega_D \Delta\Psi \,,
    \label{emf}
\end{eqnarray}
where $\Delta\Psi$ is the magnetic flux connecting the black hole to the disk, 
$\Omega_H$ is the angular velocity of the black hole
\begin{eqnarray}
    \Omega_H = \frac{a}{2M r_H} \,,
\end{eqnarray}
which is constant over the horizon of the black hole. For a thin Keplerian disk
the angular velocity of the disk is given by Eq.~(\ref{wd}).

\begin{figure}
\includegraphics[width=8.5cm]{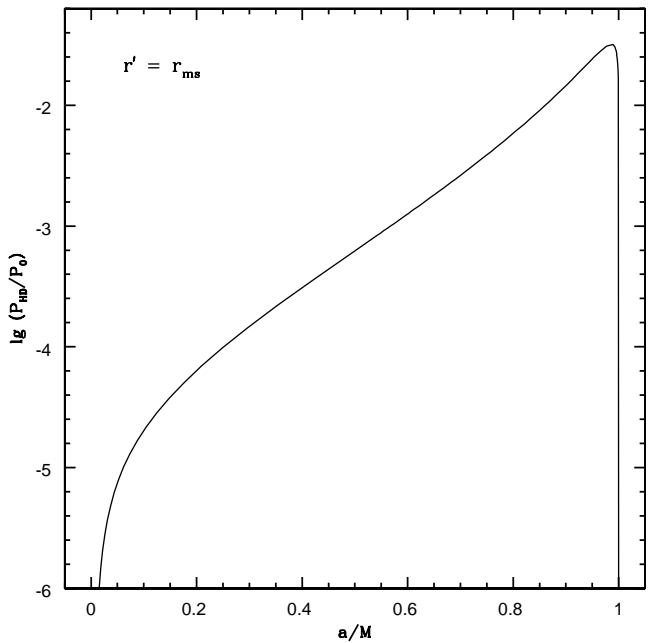}
\caption{\label{fig5} The power produced by the magnetic connection as a
function of the black hole spin: The magnetic field produced by a toroidal
electric current flowing on the inner edge of the disk connects the horizon of
the black hole to the regions of the disk with $r> r_{in}$, where $r_{in}$ ($>
r_{ms}$, where $r_{ms}$ is the radius of the marginally stable orbit where the
inner boundary of the disk is) is defined by Eq.~(\ref{rin}). The power peaks
at $a/M = 0.986$ and drops to zero as $a/M$ approaches $1$.}
\end{figure}

\begin{figure}
\includegraphics[width=8.5cm]{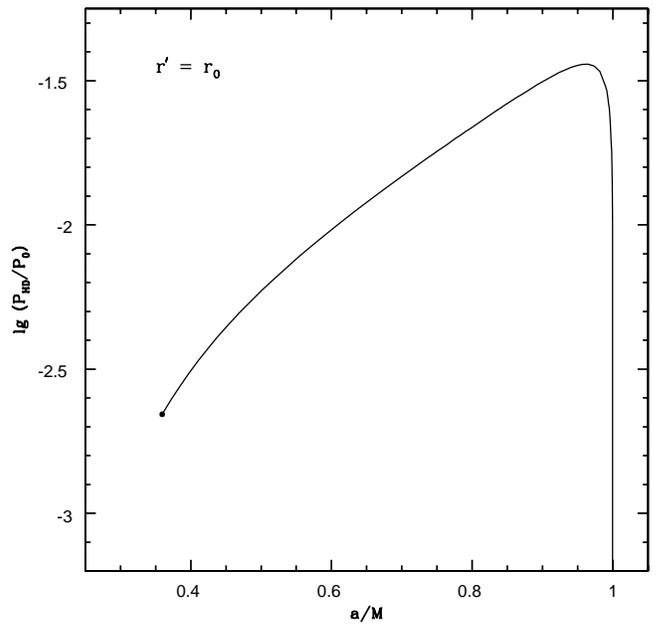}
\caption{\label{fig6} The power produced by the magnetic connection as a
function of the black hole spin: The magnetic field produced by a toroidal
electric current flowing on a circle of radius $r_0$ [defined by Eq.~(\ref{r0})]
in the equatorial plane connects the portions of the horizon of the black hole 
with $0<\theta <\pi/3$ and $2\pi/3<\theta <\pi$ to the whole disk from $r = 
r_{ms}$ to $r = \infty$. The power peaks at $a/M = 0.961$ and drops to zero as 
$a/M$ approaches $1$. The thick point is at $a/M = 0.3594$ below which steady 
state solutions do not exist.}
\end{figure}

The black hole and the disk form a closed electric circuit, an electric current
flows along the magnetic field lines connecting the black hole to the disk and 
closes itself inside the disk and the black hole. We assume that there is a thin
plasma corona around the black hole and the disk, which can carry the poloidal
current flowing between the black hole and the disk \cite{bla77}. In the
magnetosphere the resistivity along the magnetic field lines is negligible,
while the resistivity perpendicular to the magnetic field lines is large. Thus
the electric current flows along the magnetic field lines in the corona without
dissipation. Suppose the disk and the black hole rotates in the same direction,
then the EMF of the black hole, $\Delta{\cal E}_H$, and the EMF of the disk,
$\Delta{\cal E}_D$, have opposite signs. The direction of the electric current,
and thus the direction of the transfer of energy and angular momentum, is
determined by the sign of $\Delta{\cal E}_H + \Delta{\cal E}_D$. If $\Delta{\cal
E}_H + \Delta{\cal E}_D > 0$, the EMF of the black hole dominates the EMF of
the disk, so the black hole ``charges'' the disk: energy and angular momentum
are transfered from the black hole to the disk. On the other hand, if $\Delta
{\cal E}_H + \Delta{\cal E}_D < 0$, the EMF of the disk dominates the EMF of
the black hole, so the disk ``charges'' the black hole, energy and angular
momentum are transfered from the disk to the black hole \cite{li00b}.

An infinite number of adjacent infinitesimal poloidal electric current loops
flow between the black hole and the disk along the magnetic field lines
connecting them. Each infinitesimal current loop produces an infinitesimal
torque and an infinitesimal power on the disk \cite{li00a}
\begin{eqnarray}
    \Delta T_{HD} = \left(\frac{\Delta\Psi}{2\pi}\right)^2 \,\frac{\Omega_H
          - \Omega_D}{\Delta Z_H} \,,
    \label{toq1}
\end{eqnarray}
\begin{eqnarray}
    \Delta P_{HD} = \Delta T_{HD} \Omega_D \,,
    \label{pow1}
\end{eqnarray}
where $\Delta Z_H$ is the resistance of the black hole associated with the
infinitesimal current. In Eq.~(\ref{pow1}) we have used the fact that the disk
is perfectly conducting, so the magnetic field lines are frozen in and
corotate with the disk. The summation of all $\Delta T_{HD}$ gives the the
total torque on the disk, the summation of all $\Delta P_{HD}$ gives the the
total power on the disk. In the limit $\Delta r \rightarrow 0$ and $\Delta\Psi
\rightarrow 0$, the summation is replaced by integration. Then, after
taking into account that a disk has two surfaces, we obtain the total torque
and the total power on the disk
\begin{eqnarray}
    T_{HD} = 4\pi \int_{r_a}^{\infty} Hr dr \,, \hspace{0.4cm}
    P_{HD} = 4\pi \int_{r_a}^{\infty} \Omega H r dr \,,
    \label{pow2}
\end{eqnarray}
where $r_a$ is the radius in the disk where the magnetic connection starts
($r_a = r_{in}$ for $r^\prime = r_{ms}$; $r_a = r_{ms}$ for $r^\prime = r_0$),
\begin{eqnarray}
    H \equiv \frac{1}{8\pi^2 r} \left(\frac{d\Psi}{d r}\right)^2\,
         \frac{\Omega_H - \Omega_D}{- d Z_H/ dr} \,,
    \label{hh}
\end{eqnarray}
where we have treated the resistance of the black hole, $Z_H$, as a function
of the radius of the disk, which is defined by a map from the horizon of the
black hole to the surface of the disk given by the magnetic field lines
connecting them.

For a Kerr black hole we have \cite{mac82}
\begin{eqnarray}
    \frac{d Z_H}{d\theta} = \frac{R_H}{2\pi}\,\frac{r_H^2 + a^2 \cos^2\theta}
         {\left(r_H^2 + a^2\right) \sin\theta} \,,
    \label{dz1}
\end{eqnarray}
where $\theta$ is the polar angle coordinate on the horizon. Then, $dZ_H/dr$ 
can be calculated through
\begin{eqnarray}
    \frac{dZ_H}{dr} = \frac{dZ_H}{d\theta}\,\frac{d\theta}{dr} \,,
    \label{dz2}
\end{eqnarray}
where $\theta = \theta(r)$ is a map between the $\theta$ coordinate on the
horizon and the $r$ coordinate on the disk, which is induced by the magnetic
field lines connecting the disk to the horizon
\begin{eqnarray}
    \Psi\left(r_H,\theta; r^\prime\right) = \Psi\left(r, \frac{\pi}{2};
          r^\prime\right) \,.
    \label{map}
\end{eqnarray}
Since $d\theta/dr<0$ and $dZ_H/d\theta>0$, we have $dZ_H/dr<0$.

\begin{figure}
\includegraphics[width=8.5cm]{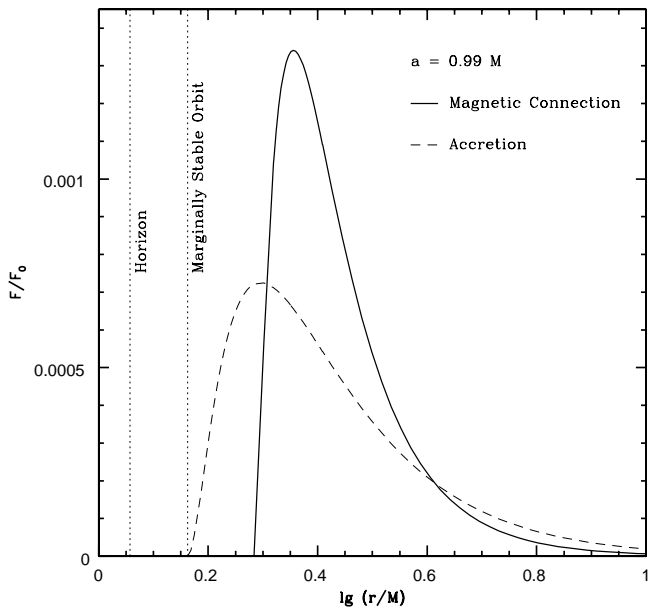}
\caption{\label{fig7} The radiation flux of a non-accretion disk magnetically
coupled to a Kerr black hole of mass $M$ and specific angular momentum $a =
0.99 M$: The poloidal magnetic field is generated by a toroidal electric
current flowing on the inner boundary of the disk which is at the marginally
stable orbit. The magnetic field connects the horizon of the black hole to the
disk in the regions with $r > r_{in}$ where $r_{in}$ is defined by Eq.
(\ref{rin}). The dashed curve is the radiation flux of a standard accretion
disk around the same black hole.}
\end{figure}

\begin{figure}
\includegraphics[width=8.5cm]{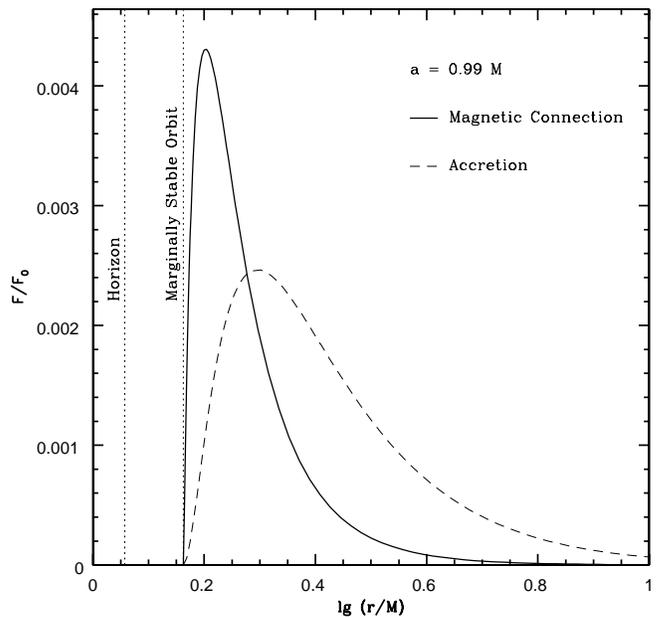}
\caption{\label{fig8}  The radiation flux of a non-accretion disk magnetically
coupled to a Kerr black hole of mass $M$ and specific angular momentum $a =
0.99 M$: The poloidal magnetic field is generated by a toroidal electric
current flowing on a circle of radius $r_0$ inside the inner boundary of the
disk, where $r_0$ is defined by Eq.~(\ref{r0}). The magnetic field connects
the portions of the horizon of the black hole with $0<\theta < \pi/3$ and $2\pi/3
< \theta < \pi$ to the disk from $r = r_{ms}$ to $r = \infty$. The dashed curve 
is the radiation flux of a standard accretion disk around the same black hole.}
\end{figure}

For a given toroidal current loop at $r = r^\prime$, we can calculate the
magnetic flux with Eqs. (\ref{flux} -- \ref{vec4}), determine the innermost
flux surface and thus $r_a$, determine the map defined by Eq.~(\ref{map}).
Then, with Eq.~(\ref{pow2}) and Eq.~(\ref{hh}), we can calculate the power of
energy transfered by the magnetic connection. Let us define a characteristic
magnetic field strength by
\begin{eqnarray}
    B_0 \equiv \frac{2 I}{M} \,,
    \label{b0}
\end{eqnarray}
then we have $\Psi = \overline{
\Psi}\, B_0 M^2$, $T_{HD} = \overline{T}\, B_0^2 M^3$, $P_{HD} = \overline{P}\,
B_0^2 M^2$, where the dimensionless quantities $\overline{\Psi}$, $\overline{T}$,
and $\overline{P}$ depend only on $a/M$ and $r^\prime/M$. We have calculated
$\overline{P}$ for two cases: (1) $r^\prime = r_{ms}$, i.e., the toroidal
current flows on the inner boundary of the disk; the radius where the innermost
flux surface touches the disk ($r_{in}$) is shown in Fig.~\ref{fig1}; (2)
$r^\prime = r_0$ where $r_0$ is defined by Eq.~(\ref{r0}), so that the magnetic
field connects the portions of the horizon of the black hole with $0<\theta <
\pi/3$ and $2\pi/3 < \theta <\pi$ to the whole disk from $r = r_{ms}$ to $r = 
\infty$; the values of $r_0/M$ as a function of $a/M$ are plotted in Fig.~\ref{fig2}. 
The results for $\overline{P} = P/P_0$, where $P_0 \equiv B_0^2 M^2$, are shown 
in Fig.~\ref{fig5} and Fig.~\ref{fig6}, respectively for the two cases.

In Fig.~\ref{fig5}, the toroidal electric current flows on the marginally
stable orbit (i.e. the inner edge of the disk). The poloidal magnetic field
generated by such a current connects the horizon of the black hole to the
regions of the disk with $r > r_{in}$, where $r_{in} > r_{ms}$ is defined by 
Eq.~(\ref{rin}). Steady state solutions exist only for $a/M \ge 0.01544$. The 
power produced by the magnetic connection peaks at $a/M = 0.986$ and drops to 
zero
as $a/M$ approaches $1$, which is a magnification of the fact that an extreme
Kerr black hole expels magnetic fields \cite{bic85}. If the black hole is spun
down from $a/M = 0.998$ to $a/M = 0.01544$, about $0.03 Mc^2$ energy is
extracted from the black hole.
In Fig.~\ref{fig6}, the toroidal electric current flows on a circle of radius
$r_0$ defined by Eq.~(\ref{r0}). The poloidal magnetic field generated
by such a current connects the portions of the horizon of the black hole with
$0<\theta < \pi/3$ and $2\pi/3 <\theta < \pi$ to the whole disk from $r =
r_{ms}$ to $r = \infty$. Steady solutions exist only for $a/M \ge 0.3594$. The
power produced by the magnetic connection peaks at $a/M = 0.961$ and drops to
zero as $a/M$ approaches $1$. If the black hole is spun down from $a/M = 0.998$
to $a/M = 0.3594$, about $0.09 Mc^2$ energy is extracted from the black hole.

%\section 4
\section{The radiation flux of the disk}
\label{sec4}

The energy pumped into the disk by the magnetic connection can be dissipated
and radiated away by the disk. In a steady state without accretion, the
radiation flux of the disk, which is the energy radiated per unit time and per
unit area by the disk as measured by an observer corotating with the disk, is
\cite{li00c}
\begin{widetext}
\begin{eqnarray}
    F = \left\{\begin{array}{ll}
       \frac{1}{r}\left(-\frac{d \Omega_D}{d r}\right) \left(E^+-\Omega_D
       L^+\right)^{-2}\int_{r_a}^r\left(E^+-\Omega_D L^+\right)Hr dr\,,
       & \mbox{if $r>r_a$} \\
       0, & \mbox{if $r<r_a$}
       \end{array}
       \right.\,,
    \label{eflux}
\end{eqnarray}
\end{widetext}
where $H$ is defined by Eq.~(\ref{hh}), $E^+$ and $L^+$ are the specific energy
and the specific angular momentum of a particle in the disk, respectively. For
a thin Keplerian disk, we have \cite{nov73,pag74}
\begin{eqnarray}
    E^+ = \frac{1 - 2 \frac{M}{r} + a \left(\frac{M}{r^3}\right)^{1/2}}{
        \sqrt{1 - 3 \frac{M}{r} + 2 a \left(\frac{M}{r^3}\right)^{1/2}}} \,,
\end{eqnarray}
\begin{eqnarray}
    L^+ = \left(M r\right)^{1/2}\,
        \frac{1 - 2 a \left(\frac{M}{r^3}\right)^{1/2} + \frac{a^2}{r^2}}
        {\sqrt{1 - 3 \frac{M}{r} + 2 a \left(\frac{M}{r^3}\right)^{1/2}}} \,,
\end{eqnarray}
and
\begin{eqnarray}
    E^+ - \Omega_D L^+ = \frac{\sqrt{1 - 3 \frac{M}{r} + 2 a \left(\frac{M}
        {r^3}\right)^{1/2}}}{1 + a \left(\frac{M}{r^3}\right)^{1/2}} \,.
    \label{ewl}
\end{eqnarray}

Since $r_{ms} \le r_a$, $F$ is always zero at the inner boundary of the disk.
As $r\rightarrow \infty$, $H$ declines quickly and the integration in 
Eq.~(\ref{eflux}) converges. Thus, as $r\rightarrow\infty$, $F$ approaches
$\propto r^{-3.5}$.

As in Sec. \ref{sec3}, we can calculate the magnetic flux with Eqs. (\ref{flux} 
-- \ref{vec4}) for a given toroidal current loop at $r = r^\prime$, then
determine $r_a$ and the map defined by Eq.~(\ref{map}). Then, with Eqs.
(\ref{hh} -- \ref{dz2}), Eq.~(\ref{eflux}), and Eq.~(\ref{ewl}), we can
calculate the radiation flux of the disk. Using the characteristic magnetic 
field strength defined by Eq.~(\ref{b0}), we have $F = \overline{F}\, F_0$
where $F_0 \equiv B_0^2$, the dimensionless quantity $\overline{F}$ depends on 
$a/M$ and $r^\prime/M$ only. We have calculated $\overline{F}$ for the 
following two cases: (1) $r^\prime = r_{ms}$, $a/M = 0.99$; (2) $r^\prime = 
r_0$, $a/M = 0.99$. The results are shown in Fig.~\ref{fig7} and Fig.~\ref{fig8}, 
respectively. For comparison, we have also shown the radiation
flux of a standard thin Keplerian disk (see Appendix \ref{app}) around the same 
Kerr black hole in Fig.~\ref{fig7} and Fig.~\ref{fig8}.

\begin{figure}
\includegraphics[width=8.5cm]{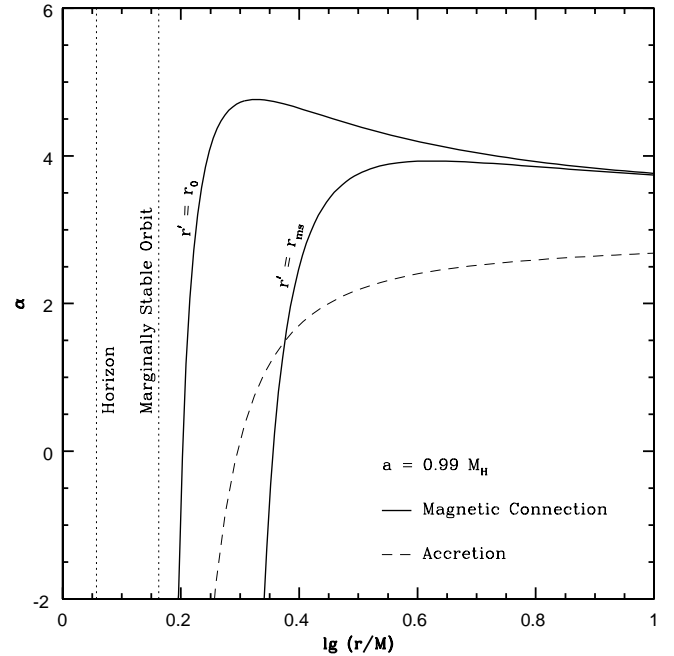}
\caption{\label{fig9} The emissivity index $\alpha \equiv - d \ln F / d \ln r$
produced by the magnetic connection for the models in Fig.~\ref{fig7} and Fig.
\ref{fig8} (labeled with ``$r^\prime = r_{ms}$'' and ``$r^\prime = r_0$''
respectively). For comparison, the emissivity index produced by the standard
accretion is shown with the dashed curve.}
\end{figure}

From Fig.~\ref{fig7} and Fig.~\ref{fig8}, we see that the radiation flux produced
by the magnetic connection is very different from that produced by the standard
accretion: They peak at different radii, and at large radii the radiation
flux produced by the magnetic connection declines faster than that produced by
the standard accretion. The position of the peak of the radiation flux produced
by the magnetic connection depends on the position of the toroidal electric
current that generates the poloidal magnetic field. To quantify the variation
speed of the radiation flux with radius, we have calculated the emissivity
defined by $\alpha \equiv - d \ln F / d \ln r$, the results are shown in 
Fig.~\ref{fig9}. We see that, the magnetic connection can produce a very large
emissivity index compared to the standard accretion, especially when the
magnetic field gets more concentrated toward the center of the disk. As
$r \rightarrow \infty$, the emissivity index of the magnetic connection
approaches $3.5$, while the emissivity index of accretion approaches $3$.

%\section 5
\section{Conclusions}
\label{sec5}
In this paper we have constructed a toy model for the magnetic connection
between a black hole and a disk. Assuming a toroidal electric current flows on
a circle in the equatorial plane of a Kerr black hole, we have calculated the
generated poloidal magnetic field which connects the horizon of the black hole
to a thin Keplerian disk around the black hole, the power produced by the
magnetic connection, and the radiation flux of the disk with the assumption
of axisymmetry and steady state. We have also compared the results with that
of a standard accretion disk.

We have particularly considered two cases: (1) The toroidal electric current
flows on the marginally stable orbit where the inner boundary of the disk is.
The magnetic field produced by such a toroidal current connects the horizon of
the black hole to the disk from $r = r_{in}$ to $r = \infty$, where $r_{in} >
r_{ms}$ is defined by Eq.~(\ref{rin}). (2) The toroidal electric current flows
on a circle of radius $r_0$ inside the marginally stable orbit, where $r_0 <
r_{ms}$ is defined by Eq.~(\ref{r0}). The magnetic field produced by such a
toroidal current connects the portions of the horizon of the black hole with
$0<\theta < \pi/3$ and $2\pi/3<\theta < \pi$ to the whole disk from $r = 
r_{ms}$ to $r = \infty$.

We see that, a magnetic connection between a black hole and a disk is naturally
produced by a toroidal current flowing around the black hole in the equatorial
plane. The magnetic connection produces a power on the disk without need of
accretion, the rotational energy of the black hole provides the energy source.
The radiation flux of the disk produced by the magnetic connection is very
different from that produced by accretion. Though the position of the peak of
the radiation flux depends on the position of the toroidal electric current, at
large radii the radiation flux of magnetic connection always declines faster
than that of accretion. Over a large range of radii the emissivity index of
magnetic connection is bigger than that of accretion. At large radii the
emissivity index of magnetic connection approaches $3.5$ from above while the 
emissivity index of accretion approaches $3$ from below.

\acknowledgments{I am very grateful to Bohdan Paczy\'nski for stimulating
discussions. This work was supported by a Harold W. Dodds Fellowship of 
Princeton University.}

\appendix

\section{Radiation Flux of a Standard Accretion Disk}
\label{app}

The radiation flux of a standard thin Keplerian disk around a 
Kerr black hole is \cite{nov73,pag74}
\begin{eqnarray}
    F_{acc} = \frac{1}{4\pi r} \dot{M}_D f \,,
    \label{facc}
\end{eqnarray}
where $\dot{M}_D$ is the mass accretion rate and
\begin{widetext}
\begin{eqnarray}
    f &=& \frac{3}{2 M}\frac{1}{x^2 \left(x^3 -3x +2s\right)}\left[x-x_0
        -\frac{3}{2} s \ln\left(\frac{x}{x_0}\right)\right.\nonumber\\
        &&- \frac{3\left(x_1-s\right)^2}
        {x_1 \left(x_1-x_2\right)\left(x_1-x_3\right)}\ln\left(\frac{x-x_1} 
	{x_0-x_1}\right) -\frac{3\left(x_2-s\right)^2}
        {x_2 \left(x_2-x_1\right)\left(x_2-x_3\right)}\ln\left(\frac{x-x_2} 
	{x_0-x_2}\right) \nonumber\\
        &&\left.- \frac{3\left(x_3-s\right)^2}
        {x_3 \left(x_3-x_2\right)\left(x_3-x_1\right)}\ln\left(\frac{x-x_3}
	{x_0-x_3}\right)\right]\,,
    \label{f0a}
\end{eqnarray}
\end{widetext}
where $s\equiv a/M$, $x\equiv (r/M)^{1/2}$, $x_0\equiv (r_{ms}/M)^{1/2}$, 
$x_1$, $x_2$, and $x_3$ are the three roots of $x^3 -3x +2s =0$
\begin{subequations}
\begin{eqnarray}
   x_1 &=& 2\cos\left(\frac{1}{3}\arccos s -\frac{\pi}{3}\right)\,,\\
   x_2 &=& 2\cos\left(\frac{1}{3}\arccos s +\frac{\pi}{3}\right)\,,\\
   x_3 &=& - 2\cos\left(\frac{1}{3}\arccos s \right)\,.
   \label{x123}
\end{eqnarray}
\end{subequations}
It can be checked that $f(r=r_{ms})=0$, and $f(r\gg r_{ms})\approx 3M/
2r^2$.

%REFERENCES


\begin{thebibliography}{99}

\bibitem{bla99} R. D. Blandford, in {\it Astrophysical Disks: An EC Summer
	School}, edited by J. A. Sellwood and J. Goodman (ASP, San Francisco,
	1999), P. 265.

\bibitem{li00a} L. -X. Li, Astrophys. Journ. {\bf 533}, L115 (2000).

\bibitem{li00b} L. -X. Li and B. Paczy\'nski, Astrophys. Journ. {\bf
        534}, L197 (2000).

\bibitem{li00c} L. -X. Li, Astrophys. Journ. {\bf 567}, 463 (2002).

\bibitem{nov73} I. D. Novikov and K. S. Thorne, in {\it Black Holes}, edited 
	by C. DeWitt and B. S. DeWitt (Gordon and Breach, New York, 1973),
        P. 343.

\bibitem{pag74} D. N. Page and K. S. Thorne, Astrophys. Journ. {\bf 191}, 499
	(1974).

\bibitem{tho74} K. S. Thorne, Astrophys. Journ. {\bf 191}, 507 (1974).

\bibitem{zna78} R. L. Znajek, Mon. Not. Roy. Ast. Soc {\bf 185}, 833 (1978).

\bibitem{dam78} T. Damour, Phys. Rev. D {\bf 18}, 3598 (1978).

\bibitem{car79} B. Carter, in {\sl General Relativity: an Einstein Centenary
        Survey}, edited by S. W. Hawking and W. Israel (Cambridge University
        Press, Cambridge, 1979), P. 294.

\bibitem{li00d} L. -X. Li, Phys. Rev. D {\bf 61}, 084016 (2000).

\bibitem{bla77} R. D. Blandford and R. L. Znajek, Mon. Not. Roy. Ast. Soc
        {\bf 179}, 433 (1977).

\bibitem{mac82} D. Macdonald and K. S. Thorne, Mon. Not. Roy. Ast. Soc.
        {\bf 198}, 345 (1982).

\bibitem{phi83} E. S. Phinney, in {\it Astrophysical Jets}, edited by
        A. Ferrari and A. G. Pacholczyk (D. Reidel Publishing Co., Dordrecht,
        1983), P. 201.

\bibitem{liv99} M. Livio, G. L. Ogilvie, and J. E. Pringle, Astrophys. Journ.
        {\bf 512}, 100 (1999).

\bibitem{mis73} C. W. Misner, K. S., Thorne, K. S., and J. A. Wheeler,
	{\it Gravitation} (W. H. Freeman, San Francisco, 1973).

\bibitem{wal84} R. M. Wald, {\it General Relativity} (The University of 
	Chicago Press, Chicago, 1984).

\bibitem{zna78a} R. L. Znajek, Mon. Not. Roy. Ast. Soc. {\bf 182}, 639 (1978).

\bibitem{lin79} B. Linet, J. Phys. A: Math. Gen. {\bf 12}, 839 (1979).

\bibitem{bic85} J. Bicak and V. Janis, Mon. Not. Roy. Ast. Soc. {\bf 212},
        899 (1985).

\end{thebibliography}
\end{document}